\documentclass[useAMS,usenatbib]{mn2e}
\usepackage{psfig}
\usepackage{hhline}
\usepackage{xspace}
\usepackage{graphicx}
\usepackage{amsmath}
\usepackage{xcolor}
\usepackage{bm}
\usepackage{times}
\usepackage{amssymb,amsmath}

\newcommand{\Hubble}{\textsl{Hubble}\xspace}

\newcommand{\Fermi}{\textsl{Fermi}\xspace}

\newcommand{\Chandra}{\textsl{Chandra}\xspace}

\newcommand{\ISIS}{\textsc{ISIS}\xspace}
\newcommand{\RN}[1]{%
  \textup{\uppercase\expandafter{\romannumeral#1}}%
}

\usepackage{times}
\usepackage{amssymb,amsmath}

\def\lsim{ \lower .75ex\hbox{$\sim$} \llap{\raise .27ex \hbox{$<$}} }
\def\gsim{ \lower .75ex \hbox{$\sim$} \llap{\raise .27ex \hbox{$>$}} }

\title[The unique case of the AGN core of M87: a misaligned low power blazar?]{The unique case of the AGN core of M87: a misaligned low power blazar?}

\author[M. Lucchini, F. Krau{\ss}, S. Markoff]
{M. Lucchini$^1$\thanks{E--mail: m.lucchini@uva.nl}, F. Krau{\ss}$^{1,2}$, S. Markoff$^{\,1,2}$\\
$^1$API -- Anton Pannekoek Institute for Astronomy, University of Amsterdam, Science Park 904, 1098 XH Amsterdam, the Netherlands\\
$^2$GRAPPA -- Gravitational and Astroparticle Physics Amsterdam, University of Amsterdam, Science Park 904, 1098 XH Amsterdam, the Netherlands\\
}

\begin{document}

\maketitle

\begin{abstract} 
M87 hosts one of the closest jetted active galactic nucleus (AGN) to Earth. Thanks to its vicinity and to the large mass of is central black hole, M87 is the only source in which the jet can be directly imaged down to near-event horizon scales with radio very large baseline interferometry (VLBI). This property makes M87 a unique source to isolate and study jet launching, acceleration and collimation. In this paper we employ a multi-zone model designed as a parametrisation of general relativistic magneto-hydrodynamics (GRMHD); for the first time we reproduce the jet's observed shape and multi-wavelength spectral energy distribution (SED) simultaneously. We find strong constraints on key physical parameters of the jet, such as the location of particle acceleration and the kinetic power. However, we under-predict the (unresolved) $\gamma$-ray flux of the source, implying that the high-energy emission does not originate in the magnetically-dominated inner jet regions. Our results have important implications both for comparisons of GRMHD simulations with observations, and for unified models of AGN classes.
\end{abstract}


%%%%%%%%% DOC

\section{Introduction}
Active galactic nuclei (AGN) are accreting super-massive black holes residing at the centre of galaxies; the gravitational energy released by accretion onto such compact objects makes them the brightest non-transient sources in the sky at all wavelengths. 

Over the years, many classes of AGN have been identified on the basis of their accretion rates, viewing angle, and presence or lack of a collimated, relativistic outflows called jets (e.g. \citealt{Antonucci93}, \citealt{Urry95}). While the basic physics of the AGN phenomenon are fairly well  understood (e.g. \citealt{Shakura73}, \citealt{Blandford77}, \citealt{Blandford79}, \citealt{Blandford82}, \citealt{Narayan94}, \citealt{Blandford99}, \citealt{Abramowicz13}), a complete picture for accretion, outflow formation and ejection, and how these are coupled is still missing. A full understanding of the energy output of AGN is necessary to quantify the impact that super-massive black holes have on their environment, which in turn is needed to correctly predict galaxy formation and evolution (e.g. \citealt{Silk98}, \citealt{Dimatteo05}, \citealt{Silk13}).

One of the most well known and remarkable AGN discovered to date is the one hosted in M87, a giant elliptical galaxy in the Virgo cluster. It hosts a remarkably massive black hole ($M_{\mathrm{bh}}= 6.5\cdot 10^{9}\,M_{\odot}$, \citealt{EHT19.6}). With this mass, the gravitational radius $R_{\mathrm{g}} = GM_{\mathrm{bh}}/c^{2} =9.7\cdot 10^{14}\,\mathrm{cm}$, making $1\,\mathrm{pc} \approx 3\cdot 10^{3}\,\mathrm{R_g}$. The source is located at a distance of $D= 16.7\pm0.6$\,Mpc, estimated thanks to the surface brightness fluctuation (SBF) method using the Hubble Space Telescope Advance Camera for Surveys Virgo Cluster Survey (ACSVCS,  \citealt{Blakeslee09}), and emits a modest bolometric luminosity of $L_{\mathrm{bol}}\approx 2.7\cdot 10^{42}\,\mathrm{erg\,s^{-1}}$, which fluctuates by about 20\% due to AGN variability between the radio and X-ray bands \citep{Prieto16}. These properties combined make M87 an excellent source to study AGN in the low-luminosity regime (LLAGN), in which the in-falling material is believed to be under-luminous (e.g. \citealt{Narayan94}, see also \citealt{Yuan14} for a recent review) and pc-scale collimated jets are more likely to be formed and launched (e.g. \citealt{Nagar05}). The viewing angle of the forward jet is estimated to be between 10 and 20 degrees (e.g. \citealt{Biretta99}, \citealt{Mertens16}, \citealt{Kim18}, \citealt{Walker18}). 

Unlike many LLAGN, the jet of M87 is easily detected on a variety of physical scales: its radiative output is believed to dominate the spectral energy distribution (SED) of the AGN core (e.g. \citealt{Nemmen14}, \citealt{Prieto16}) and the outflow extends up to kpc scales (e.g. \citealt{Biretta99}, \citealt{Owen00}, \citealt{Wilson02}). The proximity of the source allows observations mapping the jet on parsec and sub-parsec scales with an accuracy beyond that achievable for more distant sources. M87 is the only source whose jet has been resolved over multiple spatial scales, from  $\approx 10^{5}\,R_{\mathrm{g}}$, with arcsec-accuracy instruments like \Hubble and \Chandra (e.g. \citealt{Biretta99}, \citealt{Wilson02}, \citealt{Cheung07}), down to $\approx 10^{1-3}\,R_{\mathrm{g}}$ with radio VLBI at $\approx 2-86\,\mathrm{GHz}$ (e.g. \citealt{Hada11}, \citealt{Asada12}, \citealt{Nakamura13}, \cite{Hada13}, \citealt{Mertens16}, \citealt{Hada16}, \citealt{Kim18}, \citealt{Walker18}). Higher frequency VLBI observations at $230\,\mathrm{GHz}$ by \cite{Doeleman12} imply very small scales ($\approx 10\,R_{\mathrm{g}}$) for the base of the jets, and observations with the full Event Horizon Telescope (EHT) array have successfully resolved the shadow of the black hole itself \citep{EHT19.1}. The only three others sources for which a similar study of the jet collimation profile has been conducted, albeit with lower angular resolution and dynamic range in observations, are Cygnus A \citep{Boccardi16}, 3C84 \citep{Giovannini18} and NGC 4261 \citep{Nakahara18}.

This wealth of high quality, high resolution VLBI data makes M87 a unique source for isolating the physics of jets in accreting black holes. The general picture that has emerged over the years is that the jet is highly collimated and parabolic in shape up to around $10^5\,R_{\mathrm{g}}$, after which it transitions to a conical profile (\citealt{Blandford79}, \citealt{Asada12}). The inner core is likely to be magnetically dominated (\citealt{Kino14}, \citealt{Hada16}), and while in the inner pc and sub-pc scale regions only sub-luminal or mildly super-luminal speeds are observed (e.g. \citealt{Mertens16}), plasma ejected from the HST-1 knot complex (located at a de-projected distance of $\approx\,5\cdot10^{5}\,R_{\mathrm{g}}$ downstream of the core) has shown super-luminal speeds up to $6\,\mathrm{c}$ \citep{Biretta99}. Taken together, these observations imply that the jet is magnetically-dominated near the base, and accelerated up to large scales of $\approx 10^{5}\,R_{\mathrm{g}}$ by converting the initial high magnetic field into bulk kinetic energy, in agreement with GRMHD simulations (e.g. \citealt{Komissarov07}, \citealt{Chatterjee19}).

Along with extensive radio monitoring, the jet of M87 has also been studied in-depth in the high-energy regime. The X-ray emission of both the core and kpc-scale jet knots (which can be resolved by the \textit{Chandra X-ray Observatory}, hereafter \Chandra) is well reproduced by a featureless absorbed power-law; the core emission is thought to be dominated by the jet (e.g. \citealt{Wilson02}, \citealt{deJong15}, \citealt{Prieto16}) rather than the accretion flow. Remarkably, HST-1 has shown strong flaring activity in the past, even outshining the core emission (\citealt{Harris03},  \citealt{Cheung07}, \citealt{Sun18}). The source is spatially unresolved in the $\gamma$-ray band, but it has been detected both by \Fermi/LAT \citep{Abdo09} and atmospheric Cherenkov telescopes (e.g. HEGRA: \citealt{Aharonian03}, H.E.S.S: \citeyear{Aharonian06}, \citealt{Albert08}, \citealt{Abramowski12b}, \citealt{Aliu12}, VERITAS: \citealt{Acciari11}, MAGIC: \citealt{Abramowski12a}). While the \Fermi/LAT data cannot easily constrain variability, VHE observations have found variability on remarkably short timescales of a few days. The 2005/2006 VHE flare detected by HESS \citep{Aharonian06}, coincided with the period of increased activity and knot ejection in HST-1, leading \cite{Cheung07} to suggest that at least part of the high-energy emission may not originate near the black hole. Recent work by \cite{Ait18} shows that both the shape of the $\gamma$-ray spectrum and detailed analysis of the variability imply that the high energy photons are likely produced in multiple components.

Despite such complex behaviour, the overall shape of the SED has been found in the past to be consistent with a standard one-zone synchrotron self-Compton (SSC) model (e.g. \citealt{Abdo09}, \citealt{deJong15}), with the caveat that the implied bulk speed of the jet is far lower than that inferred from modelling blazar SEDs, in contrast with AGN unification models \citep{Henri06}. One possible solution to this inconsistency, which is common for single-zone models, has been proposed by \cite{Tavecchio08}, who proposed that the jet is composed of an inner, relativistic spine and of a slower moving, outer sheath. The different velocities of the two components lead to enhanced inverse-Compton emission, and the different Doppler factors of the spine and the sheath as a function of the line of sight can reconcile the differences in inferred bulk speeds for aligned and misaligned sources.

The critical drawback of both single-zone and spine/sheath models is their inability to predict both the jet's shape and/or radio emission, because in these models the synchrotron self-absorption frequency is typically $\approx 10^{11}\,\mathrm{Hz}$ (e.g. \citealt{Tavecchio98}). The aim of this paper is investigate whether this limitation also applies to in-homogeneous, multi-zone models by building on the work of \cite{Prieto16}, who fitted the radio through X-ray SED of M87 with the multi-zone \texttt{agnjet} model developed by \cite{Markoff05}. For the first time we use a semi-analytic model to reproduce both the jet shape, inferred from VLBI imaging, and the SED of an AGN jet, using the \texttt{bljet} model first presented in \cite{Me18}, hereafter Paper \RN{1}. By using both constraints at the same time, we show that we have little degeneracy in our model, and can put strong constraints on the origin of the $\gamma$-ray emission of the source. 

The paper is structured as follows: in Section 2 we build an updated multi-wavelength SED with improved X-ray coverage, in Section 3 we present the model used and apply it to the M87 core emission, in Section 4 we discuss the implications of our modelling, and in Section 5 we summarise our findings. Throughout the paper we assume a luminosity distance to the source of $16.8\,\mathrm{Mpc}$, a black hole mass of $6.5\times 10^{9}\,\mathrm{M_{\odot}}$ as in \cite{EHT19.1}, and a viewing angle of $\theta = 14^{\circ}$. At the assumed distance, an angular resolution of $0.4\arcsec$ corresponds to a physical size of $\approx 35\,\mathrm{pc}$. 

%-----------------------------------------------------------------
\begin{figure*}
\caption{Sky map of \Chandra images of M87 in J2000.0 coordinates. The left panel shows a wide view of the jet of M87 and surrounding gas in the host galaxy. The extraction region of gas south of the jet is shown in red. The middle panel shows a closeup of the HRC image (Obs ID 13515) of M87, where the HST-1 and the core of M87 are clearly visible. The extraction region for the extended jet is shown in white. The section that is shown in the middle panel is shown with a gray box in the left panel. The right panel shows an ACIS image of M87 (observation 13964), as well as the observation regions for the core and HST-1, which are more difficult to separate. The section of the image is shown as a gray box without connecting lines in the left panel. 
}
\includegraphics[scale=0.5]{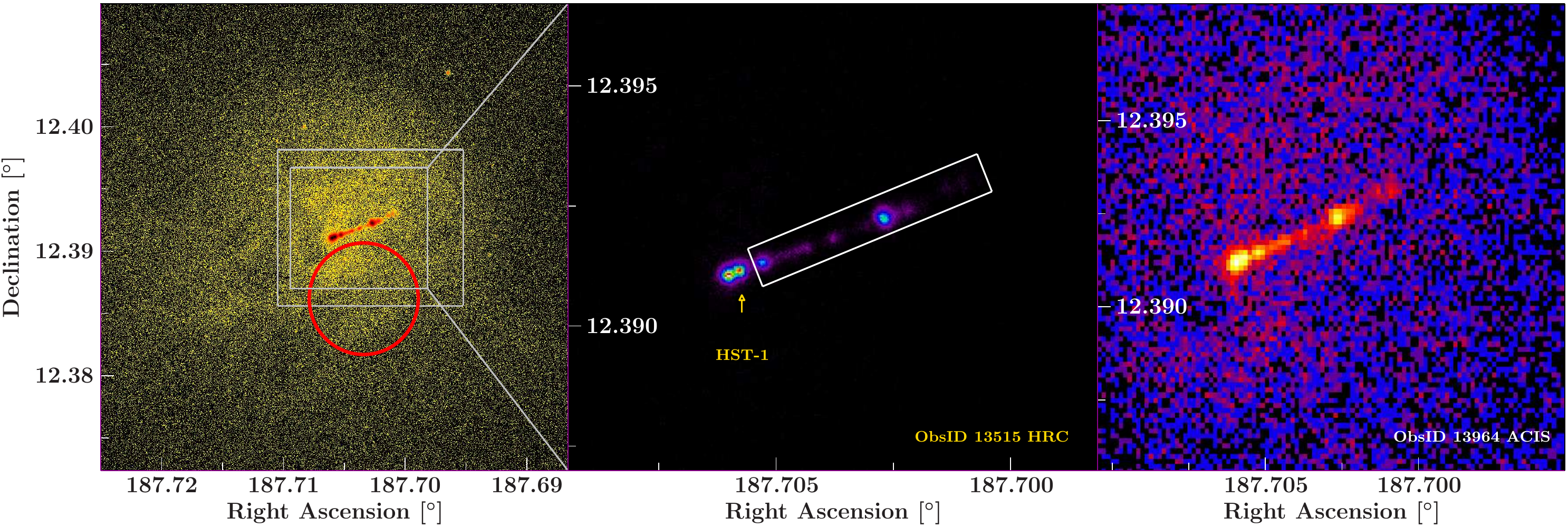}
\label{fig-chandrasky}
\end{figure*}
% ---------------------------------------------------

%-----------------------------------------------------------------
\begin{table*}
\caption{List of the regions used for the \Chandra data extraction for each observation. $\alpha$ and $\delta$ specify the right ascension and declination, respectively, in the J2000.0 system. The variables $r$, $\Theta$, $l$ and $w$, give the radius, the angle, the length and width of the observation, respectively. An example of the regions is also shown in Fig.~\ref{fig-chandrasky}. To estimate the contribution of the diffuse emission, we use three regions in all observations with $\alpha_\mathrm{component1}=187.706867^\circ$,$\delta_\mathrm{component1}$, $r_\mathrm{component1}=9.366^{\prime\prime}$; $\alpha_\mathrm{component2}=187.708754^\circ$, $\delta_\mathrm{component2}=12.389499^\circ$, $r_\mathrm{component2}=4.745^{\prime\prime}$; $\alpha_\mathrm{component3}=187.703400^\circ$, $\delta_\mathrm{component3}=12.386178^\circ$, and $r_\mathrm{component3}=16.135^{\prime\prime}$. The background is given by $\alpha_\mathrm{bkg} = 187.696525^\circ$, $\delta_\mathrm{bkg} = 12.3799006^\circ$, with a radius of $r_\mathrm{bkg}=22.487^{\prime\prime}$.}
\begin{tabular}{l|llllll}
Obs ID & $\alpha_{\mathrm{Core}}$ [$^\circ$] & $\delta_\mathrm{Core}$ [$^\circ$] & 
$r_\mathrm{Core}$ [$^{\prime\prime}$] & $\alpha_\mathrm{HST-1}$ [$^\circ$] &
$\delta_\mathrm{HST-1}$ [$^\circ$] & $r_\mathrm{HST-1}$ [$^{\prime\prime}$] \\
\hline 
13964 & 187.705896 & 12.391174 & 0.516 & 187.70566 & 12.39133  & 0.516 \\
13965 & 187.705929 & 12.391056 & 0.516 & 187.705646  & 12.391179 & 0.516  \\
14973 & 187.706058 & 12.3910247 & 0.516 & 187.705779 & 12.391179 & 0.516 \\
14974 & 187.706050 & 12.391067 & 0.516 & 187.705638  & 12.391254 & 0.516 \\
\end{tabular}
\begin{tabular}{l|lllll}
\hline
Obs ID & $\alpha_{\mathrm{jet}}$ [$^\circ$] & $\delta_{\mathrm{jet}}$ [$^\circ$] &
$\Theta_{\mathrm{jet}}$  [$^\circ$] & $l_\mathrm{jet}$ [$^{\prime\prime}$] &
$w_\mathrm{jet}$ [$^{\prime\prime}$] \\
\hline
13964 & 187.70296 & 12.392370 & 22.0006  & 18.83 & 2.951 \\
13965 & 187.70296 & 12.392370 & 22.0006 & 18.83  & 2.951 \\
14973 & 187.702971 & 12.392234 & 22.0006 & 18.963 & 3.046 \\
14974 & 187.702871 & 12.392298 & 22.0006 & 18.963 & 3.046\\
\end{tabular}
\label{tab-chandra}
\end{table*}
%-----------------------------------------------------------------

\section{Data analysis}
We compile a new multi-wavelength SED of M87 by complementing the quiescent state, $0.4\arcsec$ data at radio, sub/mm, infra-red and optical frequencies of \citealt{Prieto16} (in which the details of the data selection and reduction are reported) with additional X-ray and $\gamma$-ray coverage.

We looked for \Chandra observations coinciding as closely as possible (within a period of a few months) with the ALMA observations of June 2012. We also take the \Fermi/LAT $\gamma$-ray spectrum from the 3FGL catalogue \citep{Acero15} as representative of the source's steady-state high energy emission. %, and re-analysed the \Nu data presented by \cite{Wong17}.

\subsection{\Chandra data reduction}

The following \Chandra observations were available between December 2011 and March 2013: 13964, 13965, 13515, 14973, and 14974.

The \Chandra observations of M87 were extracted with CIAO\,4.9 using the standard pipelines. First the CIAO script chandra\_repro was run to update calibrations. As a second step, ACIS observations were extracted using specextract and ds9-generated regions. One observation, ID 13515 was taken with the High Resolution Camera (HRC),
and no spectra was extracted for this source. Specextract was run with the psf correction, to correct the ARF for the small (sub-PSF) extraction region. We extracted spectra for the core region, HST-1, the kpc-scale jet, as well as a background region.
Finally, we also extracted a spectrum to the south of the jet in order to get a spectrum of the gas in the host galaxy surrounding the jet. The gas spectrum is used as background for the spectra of the AGN components (core, HST-1 and kpc-scale jet). 

All extraction regions are reported in Table \ref{tab-chandra} and shown in Fig.~\ref{fig-chandrasky}. 
% ---------------------------------------------------
\begin{table*}
\hspace*{-0.75cm}
\caption{Best-fit parameters for the three AGN components (core, HST-1, kpc-scale jet) for each observation, listed in chronological order, as well as for the stacked spectra (obs IDs 13964, 13965, 14973). The normalisation of the power-laws is given in units of $10^{-4}$ photons/cm$^{2}$/s/keV at 1 keV. The normalisation of the core and HST-1 spectra from observation ID 14974 are clearly inconsistent with the remaining observations.}
\begin{tabular}{| l | c | c | c | c | c | c | c | c | c |}
\hline
Component & Norm & $\Gamma$ & Component & Norm & $\Gamma$ & Component & Norm & $\Gamma$ & $N_{\mathrm{h}}$\\
 & $10^{-4}$ & & & $10^{-4}$ & & & $10^{-4}$ & & $10^{20}\,\mathrm{cm^{-2}}$\\
\hline
Core, 13964 & $5.79^{+0.33}_{-0.23}$ & $2.13^{+0.08}_{-0.05}$ 
& HST-1, 13964 & $3.13^{+0.20}_{-0.20}$ & $2.64^{+0.10}_{-0.09}$
& Jet, 13964 & $6.75^{+0.29}_{-0.24}$ & $2.54^{+0.07}_{-0.06}$ & $6.9^{+1.2}_{-0.6}$\\
\hline
Core, 13965 & $5.23^{+0.24}_{-0.23}$ & $2.12^{+0.07}_{-0.06}$
& HST-1, 13965 & $2.79^{+0.17}_{-0.15}$ & $2.54^{+0.11}_{-0.08}$
& Jet, 13965 & $6.40^{+0.32}_{-0.22}$ & $2.50^{+0.08}_{-0.07}$ & $6.9^{+1.2}_{-0.6}$\\
\hline
Core, 14974 & $2.62^{+0.16}_{-0.16}$ & $1.98^{+0.10}_{-0.07}$
& HST-1, 14974 & $1.85^{+0.16}_{-0.12}$ & $2.64^{+0.15}_{-0.10}$
& Jet, 14974 & $6.34^{+0.30}_{-0.25}$ & $2.45^{+0.10}_{-0.06}$ & $6.9^{+1.2}_{-0.6}$\\
\hline
Core, 14973 & $4.56^{+0.23}_{-0.18}$ & $2.06^{+0.07}_{-0.07}$
& HST-1, 14973 & $2.78^{+0.17}_{-0.18}$ & $2.57^{+0.12}_{-0.09}$
& Jet, 14973 & $6.92^{+0.32}_{-0.26}$ & $2.57^{+0.06}_{-0.09}$ & $6.9^{+1.2}_{-0.6}$\\
\hline
Core, stacked & $5.10^{+0.27}_{-0.18}$ & $2.04^{+0.06}_{-0.03}$
& HST-1, stacked & $2.79^{+0.14}_{-0.14}$ & $2.52^{+0.08}_{-0.05}$
& Jet, stacked & $6.58^{+0.20}_{-0.25}$ & $2.50^{+0.58}_{-0.05}$ & $5.6^{+1.2}_{-1.0}$\\
\hline
\end{tabular}
\label{tab-pls}
\end{table*}
% ---------------------------------------------------

\subsection{X-ray spectral modelling}

We fit both phenomenological and physical models to the data using the Interactive Spectral Interpretation System (\ISIS) software package \citep{Houck00}, version 1.6.2-35, which enables the statistical modelling of multi-wavelength spectra using custom models. All models are folded through the detector response matrices of X-ray satellites; at all other wavelengths, the instrument response is assumed to be an identity matrix, which represents the response of a detector with effective area $= 1\,\mathrm{m^{2}}$. \Chandra spectra are binned to a signal to noise ratio of 4.5 in order to be able to use $\chi^{2}$ statistics when fitting. Each fit is performed by running the \texttt{subplex} $\chi^{2}$ minimisation algorithm, after which we refine the fit and explore parameter space by using the \ISIS implementation of a Markov Chain Monte Carlo (MCMC) routine, based on the \texttt{emcee} developed by \cite{Foreman13}. The routine initialises an ensemble of walkers (we use 100 for each free parameter) which at each iteration move through the parameter space; depending on the $\chi^{2}$ values in the new and old position the move may be accepted or rejected. We evolve the chain for 5000 iterations and discard the first 1500 as the ``burn-in'' period of the chain. In this way, the MCMC routine identifies the global minimum in the parameter space, along with possible degeneracies among parameters. The final distribution of walkers allows us to estimate the best fit values of the global minimum, and uncertainties of the fitted parameters. These are defined respectively as the peaks in the posterior distribution of the walkers, and as the intervals in the the one-dimensional histograms containing 68\% of the walkers from the end of the burn-in period to the end of the \texttt{emcee} run. We adopt the abundances of \cite{Wilms00} and set the photo-ionisation cross-sections according to \cite{Verner96}.

% ---------------------------------------------------
\begin{figure}
\hspace{-0.5cm}
\includegraphics[scale=0.62]{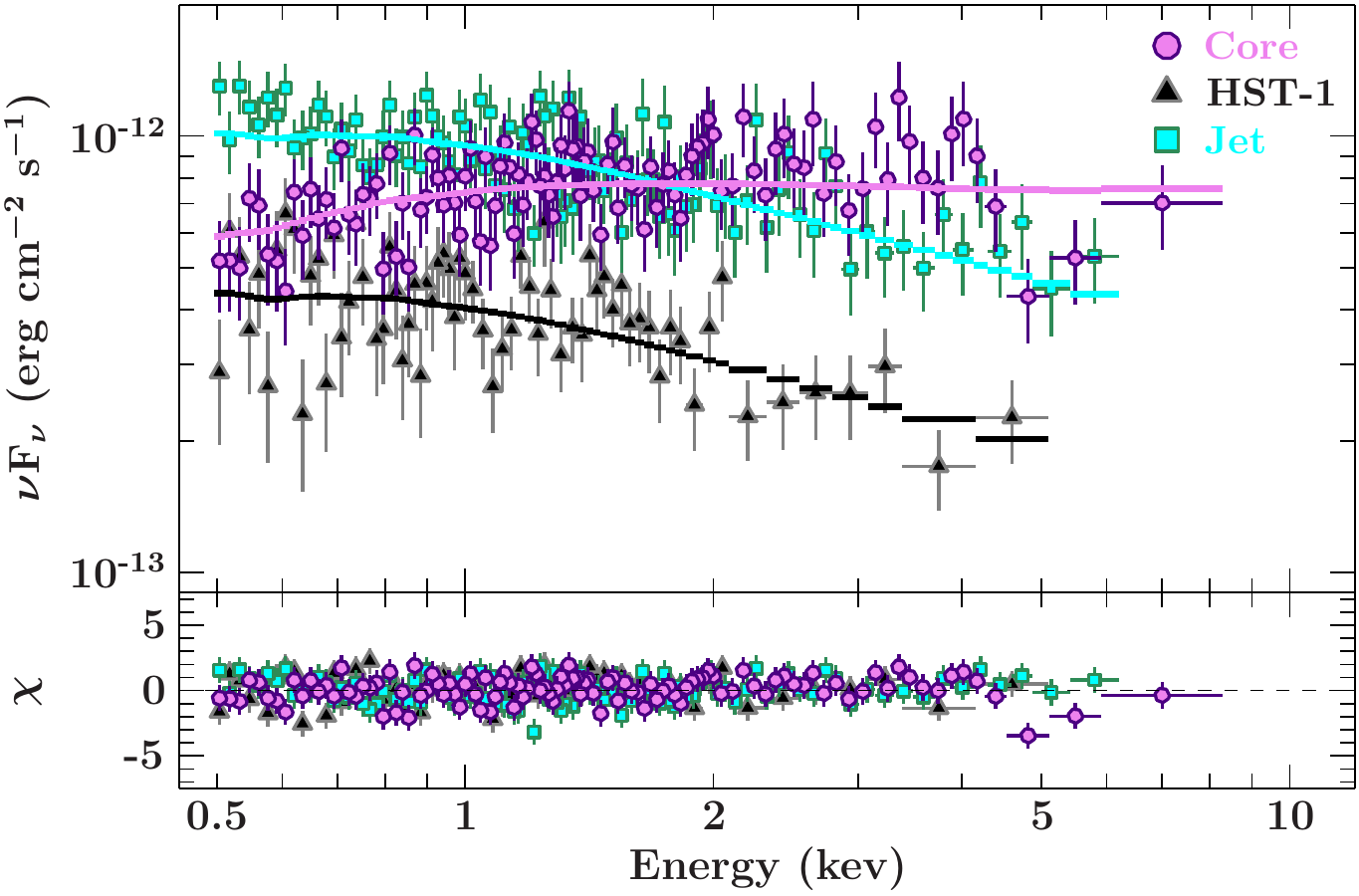}
\caption{
Combined X-ray spectra of M87. All three spectra are well fit by an absorbed power-law model. The core spectrum is harder than both the kpc-scale jet and HST-1.
}
\label{fig-Xray}
\end{figure}

We first fit the \Chandra spectra for the three components (core, HST-1 and kpc-scale jet) to ensure they are consistent with each other. Each spectrum was fitted with an absorbed power-law (\texttt{tbnew$\times$powerlaw}); in all cases the power-law model is in excellent agreement with the data ($\chi^{2}_{\mathrm{red}} = 1.05$ for the combined data set). The fits did not show any statistical improvement if we let the column density vary between spectra, so we tied $N_{\mathrm{H}}$ across the entire data set; our best-fit values show a small excess (by a factor of about 3) above the Galactic value of $1.94\times 10^{20}\,$cm$^{2}$. The best-fit values are shown in table \ref{tab-pls}, and the spectra and residuals are shown in figure \ref{fig-Xray}.

The spectra at all epochs are consistent with each other with the exception of observation ID 14974. In this epoch we find that the flux of both the core and HST-1 is lower by a factor of $\approx$ 2, while the spectral indices and kpc-scale jet remain unchanged. We do not believe this to be a physical change caused by the source's variability. This is because, in order for the variability to be physical, both the core and HST-1 would have to vary by the same amount over the same period, which is extremely unlikely. Instead, the discrepancy in flux measurements is likely caused by the difficulty in separating the core and HST-1 components in ACIS images, as shown in the right panel of figure \ref{fig-chandrasky}. Because of these systematics, we neglect the core and HST-1 spectra from observation ID 14974 in the following analysis.

% ---------------------------------------------------
\begin{table}
\hspace{-0.5cm}
\caption{List of model parameters in \texttt{bljet}; the first group of 6 is constrained by the VLBI data and fixed during spectral fitting, the second group of 5 are kept as free parameters during spectral fitting, and the last 3 are set to unity, as leaving them free did not improve the quality of the fits.}
\begin{tabular}{@{}cp{5.2cm}}
\hline
Parameter & Description \\
\hline
$r_{\mathrm{0}} = 3\,R_{\mathrm{g}}$ & The initial radius of the jet nozzle/corona; we assume the aspect ratio is $h=2r_{\mathrm{0}}$\\
$z_{\mathrm{acc}} = 2.5\cdot 10^{5}\,R_{\mathrm{g}}$ & The location where bulk acceleration of the flow stops and the jet transitions from parabolic to conical\\
$z_{\mathrm{max}}=3\cdot10^5\,\mathrm{R_{g}}$ & The total length of the jet up to which the emission is calculated\\
$\Gamma_{\mathrm{acc}}$ = 15 & The final Lorentz factor of the jet at $z_{\mathrm{acc}}$\\
$\alpha = 0.5$ & The scaling factor of the bulk Lorentz factor with distance, $\Gamma(z) \propto z^{\alpha}$\\
$\rho = 0.18$ & The collimation profile of the jet, $\theta(z)=\rho/\Gamma(z)$\\
\hline
$N_{\mathrm{j}}$ & Power channelled into the base of the jet in Eddington units\\
$\gamma_{\mathrm{e}}$ & Peak Lorentz factor of the relativistic Maxwellian distribution of electrons\\
$z_{\mathrm{diss}}$ & Distance along the jet after where particle acceleration begins\\
$p$ & Slope of the accelerated particle power-law distribution beyond $z_{\mathrm{diss}}$\\
$f_{\mathrm{sc}}$ & Particle acceleration efficiency scaling, which sets the maximum lepton energy in the power-law \\
\hline
$\sigma_{\mathrm{acc}} = 1$ & The magnetization of the jet at the end of the parabolic acceleration region.\\
$f_{\mathrm{heat}} = 1$ & The amount of heating received by the electrons at the dissipation region, which sets the minimum Lorentz factor $\gamma_{\mathrm{min}}$ of the power-law distribution. \\
$f_{\mathrm{b}} = 1$ & A dimensionless parameter responsible for setting the importance of adiabatic losses with respect to radiative ones, thus shifting the cooling break Lorentz factor in the non-thermal lepton distribution $\gamma_{\mathrm{break}}$.\\
\hline
\end{tabular}
\label{tab-bljetpars}
\end{table}
% ---------------------------------------------------

% ---------------------------------------------------
\begin{figure*}
\caption{
Jet profile in \texttt{bljet} for different parameters compared with VLBI data in the inner parabolic region. Left panel: jet acceleration profile ($\Gamma(z)\propto z^{\alpha}$); middle panel: jet terminal Lorentz factor; right panel: jet opening angle ($\theta(z) = \rho/\Gamma(z)$). The model is in good agreement with the data by taking $\alpha = 0.5$, $\Gamma_{\mathrm{acc}} = 15$ and $\rho = 0.18$, with the exception of the inner $\approx 500\,{\mathrm{R_g}}$.
}
\label{fig-jetshapes}
\hspace{-1.cm}
\includegraphics[width=\textwidth]{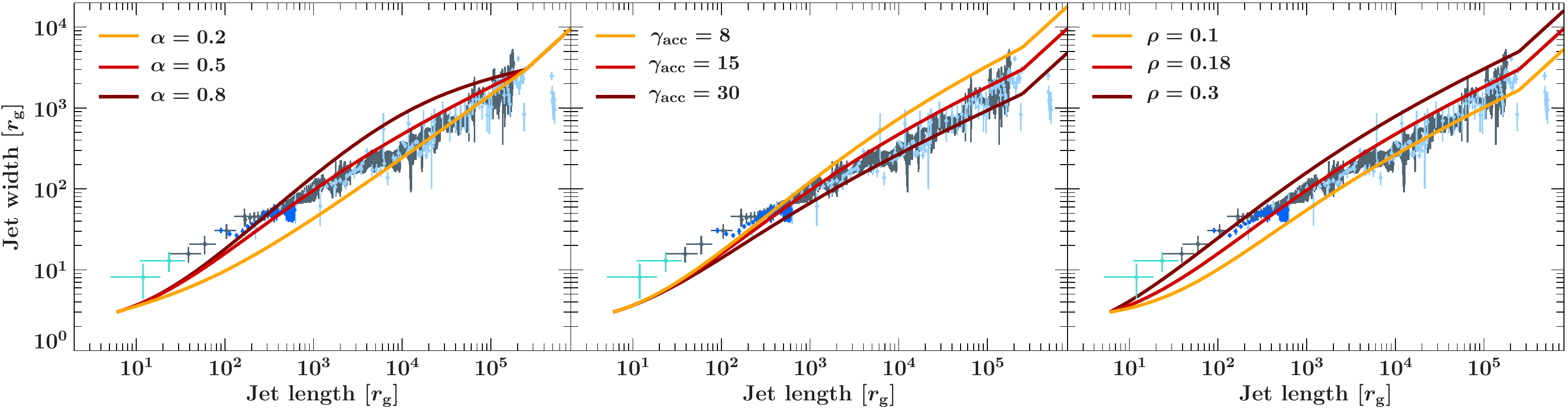}
\end{figure*}
% ---------------------------------------------------

% ---------------------------------------------------
\hspace{-0.5cm}
\begin{figure}
\includegraphics[width=\linewidth]{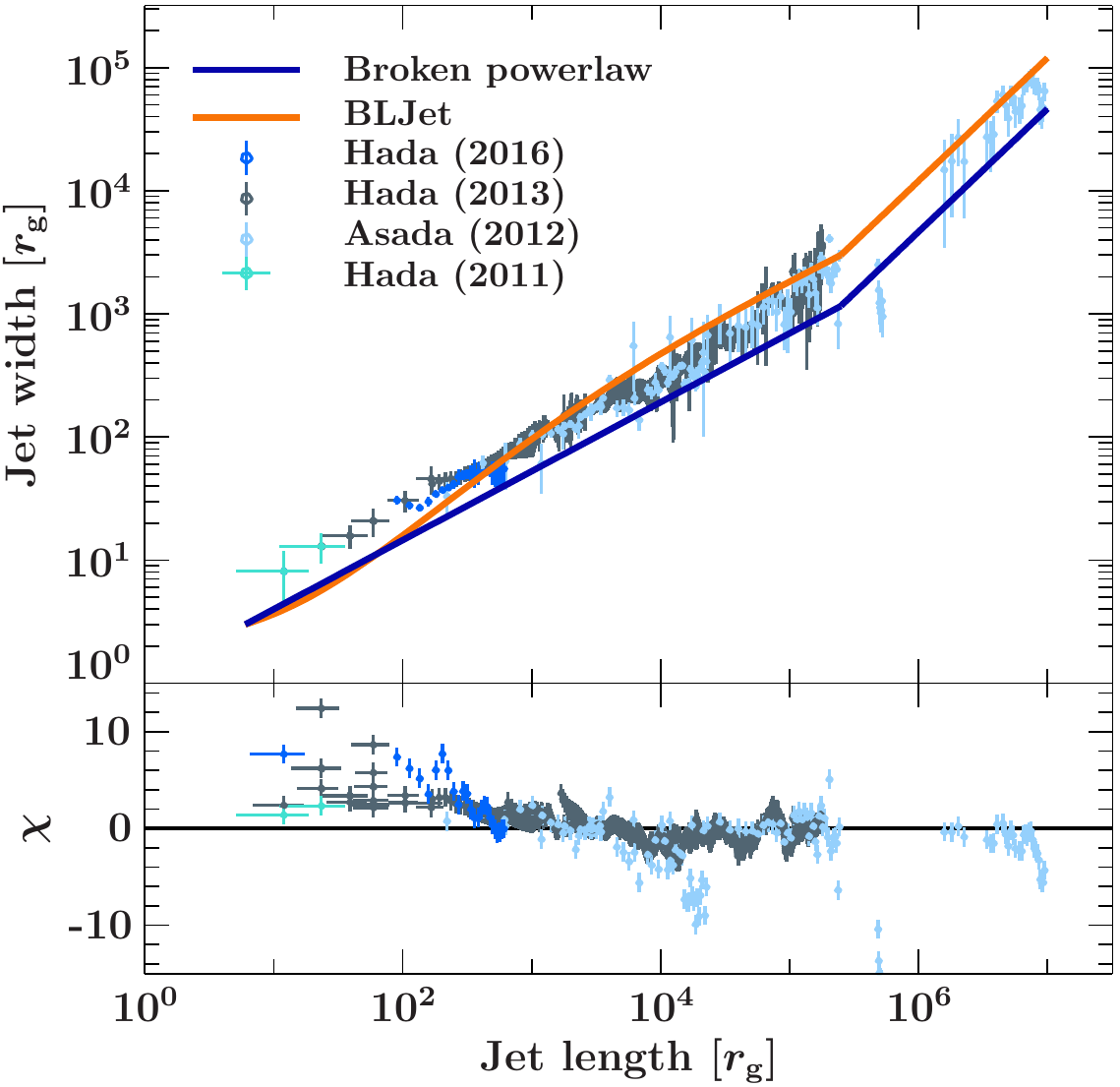}
\caption{
Top panel: jet profiles overlaid on top of the VLBI imaging data. Orange shows the jet profile assumed during the spectral fits with \texttt{bljet}; dark blue shows a broken power-law fit of the VLBI data ($r \propto z^{0.56}$ in the inner region, $r \propto z$ in the outer region). Bottom panel: model residuals for the jet geometry in \texttt{bljet}.
\label{fig-jetdata} 
}
\end{figure}
% ---------------------------------------------------

\section{Modelling the M87 core emission}

In this section we model the core SED with the \texttt{bljet} leptonic multi-zone model; the full details are contained in Paper \RN{1}, and references therein. Briefly, the model assumes that a fraction $N_{\mathrm{j}}$ of the black hole's Eddington power is injected in a highly magnetized nozzle of radius $r_{\mathrm{0}}$ and aspect ratio $h=2r_{\mathrm{0}}$, which can be thought of as related to a magnetized corona or wind (see, e.g., \citealt{Markoff05}). We define the initial magnetization as $\sigma_0 = (U_{\mathrm{b,0}} + P_{\mathrm{b,0}})/(U_{\mathrm{p,0}} + U_{\mathrm{e,0}} + P_{\mathrm{e,0}}) \gg 1$, where $U_{\mathrm{b,0}}$, $U_{\mathrm{e,0}}$, $U_{\mathrm{p,0}}$ are the initial energy densities of the magnetic field, electrons and protons in the jet, $P_{\mathrm{b,0}}$ and $P_{\mathrm{e,0}}$ the pressure of the magnetic field and electrons, and the protons are assumed to be non-relativistic and thus have negligible pressure. The jet accelerates up to a terminal Lorentz factor $\Gamma_{\mathrm{acc}}$ up to a distance $z_{\mathrm{acc}}$ by converting the initial magnetic field into bulk kinetic energy until the outflow becomes matter-dominated ($\sigma_{\mathrm{acc}} \leq 1$). The bulk Lorentz factor of the outflow is assumed to scale with distance from the black hole $z$ as a power-law: $\Gamma(z)\propto z^{\alpha}$, with $\alpha \approx 0.5$. The jet opening angle is inversely proportional to the Lorentz factor: $\theta(z) = \rho/\Gamma(z)$, where $\rho$ is a proportionality constant taken to be less than 1. The resulting jet profile is roughly parabolic in shape in the bulk acceleration region, and conical in the outer region. 

The leptons in the jet are assumed to be thermalised and relativistic up to a distance $z_{\mathrm{diss}}$ from the base (which we took to be equal to $z_{\mathrm{acc}}$ in paper \RN{1}, though this need not be the case), at which point $10\%$ of the particles are injected in a power-law tail with slope $p$. At the dissipation region the particle distribution is also assumed to be heated, parametrised by increasing the temperature of the relativistic Maxwellian by a fixed factor $f_{\mathrm{heat}}$. The energy of the cooling break in the non-thermal particle distribution is controlled by the free parameter $f_{\mathrm{b}}$, which regulates the importance of adiabatic losses with respect to radiative losses. The maximum energy reached by the particles is parametrised by the dimensionless parameter $f_{\mathrm{sc}}$ which sets the time-scale of the acceleration mechanism. In preliminary fits we found that the magnetization at the acceleration region $\sigma_{\mathrm{acc}}$ has a negligible effect on the SED as the bulk of the emission is generated fairly close to the jet base ($z\leq 10^{4}\,{\mathrm{R_g}}$), in highly magnetized regions where $\sigma \gg 1$, and therefore we fix it to $\sigma_{\mathrm{acc}} = 1$. Unlike in paper \RN{1} we also found that the SED was well matched by assuming an isothermal jet with constant temperature $\gamma_{\mathrm{e}}$, and that the break energy of the particles was consistent with the value calculated from equating adiabatic and synchrotron time-scales: $E_{\mathrm{br}}(z) = (3\beta(z) m_{\mathrm{e}}^{2}c^{4})/(4 r(z)\sigma_{\mathrm{T}}U_{\mathrm{b}}(z))$, where $z$ is the distance along the jet, $\beta(z)$ is the speed of the jet in units of $c$, $m_{\mathrm{e}}$ is the mass of the electron, $r(z)$ is the radius of the jet, $\sigma_{\mathrm{T}}$ is the Thomson cross section, and $U_{\mathrm{b}}(z)$ the magnetic field energy density along the jet. We therefore fixed both $f_{\mathrm{heat}}$ and $f_{\mathrm{b}}$ to unity.

Finally, we have expanded the inverse Compton calculation in the code to include the host galaxy's stellar photon field. Following e.g. \cite{Stawarz06}, we assume that the radiation energy density in the galaxy core is with $U_{\mathrm{rad}} = 10^{-9}\,$erg\,cm$^{-3}$ and peak temperature $T=3200\,\mathrm{K}$. The main parameters of the model are summarised in Table \ref{tab-bljetpars}.

\subsection{Matching the jet collimation profile}

Before performing spectral fits with \texttt{bljet} we match the jet geometry to the available VLBI observations of the source. The quality of the imaging data limits the allowed range of the parameters of \texttt{bljet}, particularly $\alpha$, $\Gamma_{\mathrm{acc}}$, and $\rho$. In paper \RN{1}  we took fixed values for these parameters; in this section we will show that these values allow the shape predicted by the model to match the observation fairly well. We stress that because our collimation profile model is relatively simple, our goal is to find a set of parameters that qualitatively produces a jet similar to that of M87 fit, rather than perform a quantitative statistical fit of the jet's collimation and acceleration.

We first combine the imaging observational constraints by fixing the radius of the jet nozzle $r_0$ to $3\,R_{\mathrm{g}}$, which ensures that the computed jet width never exceeds the observed value, and the distance of the bulk collimation/acceleration region $z_{\mathrm{acc}}$ to $2.5 \times 10^{5}\,R_{\mathrm{g}}$ (e.g. \citealt{Asada12}, \citealt{Hada13}), which fixes the location of the parabolic to conical transition in the jet profile. We then vary the values of $\alpha$, $\rho$ and $\Gamma_{\mathrm{acc}}$ and check their impact on the predicted jet shape. The resulting profiles are shown in figure \ref{fig-jetshapes}. In general highly collimated jets (corresponding to low values of $\alpha$ and $\rho$) predict a jet that is too narrow near the base; less collimated shapes (large values of $\alpha$ and $\rho$) instead over-predict the jet width on larger scales. Similarly, fast jets are narrower than slow jets.

As shown in figure \ref{fig-jetdata} taking $\alpha=0.5$, $\rho=0.18$ and $\Gamma_{\mathrm{acc}}=15$ provides a reasonable agreement with the imaging data, with the exception of the inner $\approx 500\,R_{\mathrm{g}}$; we thus keep these values unchanged and assume a constant geometry throughout the spectral fitting procedure.  In principle an even better agreement with the imaging data could be obtained by assuming, for example, that $\rho$ also changes with distance (the right panel of \ref{fig-jetshapes} for instance suggests that $\rho$ is around 0.3 at the base, decreasing to 0.18 further out). Imposing a wider collimation profile in the initial segments of the jet would lower the number density in these regions, while the strength of the magnetic field and total number of particles would be unchanged. As a result, the synchrotron emission in the final SED would be unchanged, but the inverse-Compton component would be suppressed slightly. As we will discuss in the next section, we don't expect the inverse Compton emission to contribute meaningfully to the core's emission from radio to X-rays, and thus our conclusions are unaffected by the mismatch in the inner jet collimation profile. 

% ---------------------------------------------------
\begin{table}
\begin{center}
\caption{Best-fit synchrotron-dominated parameters and relative uncertainties.}
\begin{tabular}{| l | c | c | c |}
\hline
Injected power $N_{\mathrm{j}}$ ($10^{-5}\,L_{\mathrm{Edd}}$) & & & $3.2^{+0.5}_{-0.3}$ \\
Dissipation distance $z_{\mathrm{diss}}$ ($\mathrm{r_{g}}$) & & & $97^{+13}_{-3}$ \\
Electron temperature $T_{\mathrm{e}}$ ($\gamma_{\mathrm{e}}$) & & & $4.2^{+1.0}_{-0.3}$ \\
Non-thermal power-law slope $p$ & & & $2.34^{+0.01}_{-0.01}$ \\
Acceleration efficiency $f_{\mathrm{sc}}$ & & & $1.^{*}_{-0.2}$ \\
\hline
BB Normalisation ($10^{-5}\,L_{39}/D^{2}_{10\,\mathrm{kpc}}$) & & & $5.1^{+0.7}_{-0.6}$ \\
BB Temperature (ev) & & & $0.28^{+0.03}_{-0.02}$ \\
\hline
$^*$: best fit consistent with allowed upper limit
\end{tabular}
\label{tab-bljetfitpars}
\end{center}
\end{table}
% ---------------------------------------------------

% ---------------------------------------------------
\begin{figure}
%\hspace{-0.5cm}
\includegraphics[scale=0.6]{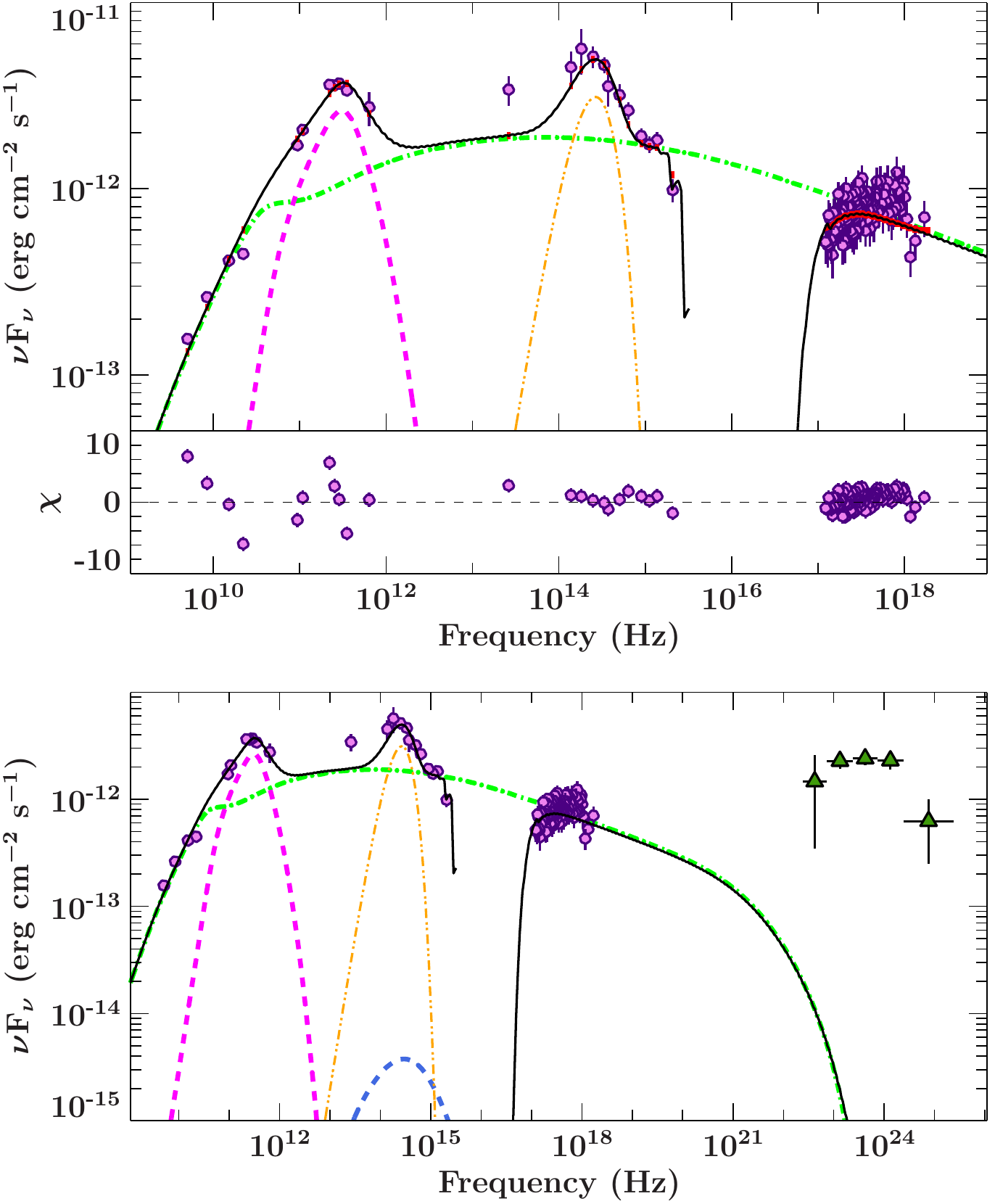}
\caption{
Top panel: radio through X-ray sub-arcsec core SED of M87 with our best synchrotron-dominated fit. The dashed pink line represents thermal synchrotron emission from the jet base/corona; the dot-dashed green line de-absorbed, non-thermal synchrotron emission, the double dot-dashed yellow line the black body contribution, approximating the host galaxy contribution. The black solid line shows the total model flux and includes the effects of absorption. 
Bottom panel: radio through $\gamma$-ray SED of M87. The inverse-Compton contribution of the jet core is shown by the dashed blue line.
\label{fig-sedbljet}
}
\end{figure}
% ---------------------------------------------------

\subsection{Broadband spectral modelling: a synchrotron-dominated inner core}

The constraints imposed by the imaging data on the jet shape leave only 5 of the fitted parameters in \texttt{bljet} to be free: the injected jet power $N_{\mathrm{j}}$, location of the particle acceleration region $z_{\mathrm{diss}}$, electron temperature $\gamma_{\mathrm{e}}$, slope of the non-thermal distribution $p$, and particle acceleration efficiency $f_{\mathrm{sc}}$.

We find that the data requires an excess in the optical/near-IR bands, which we model as a single black body. The model syntax assigned in \ISIS to the the full SED is \texttt{tbnew$\times$(bljet+bbody)}.

Our best fit of the SED is shown in figure \ref{fig-sedbljet}, and the best-fit parameters and uncertainties are reported in table \ref{tab-bljetfitpars}. The radio and X-ray emission is mainly due to non-thermal synchrotron, while the ALMA band is dominated by thermal synchrotron emission from the jet nozzle and the optical/IR emission shows a prominent thermal bump. The model is in excellent agreement with the data up through the X-ray band; furthermore after running \texttt{emcee} we do not find any significant degeneracy in any of the free parameters, which are all very well constrained. However, the model predicts a $\gamma$-ray flux of $\approx 10^{-14}\,$erg s$^{-1}$ cm$^{-2}$, far below the \Fermi/LAT spectrum; we discuss the implications of this finding in section 4.

\subsection{Broadband spectral modelling: SSC-dominated regime}

\begin{figure}
%\hspace{-0.5cm}
\includegraphics[scale=0.6]{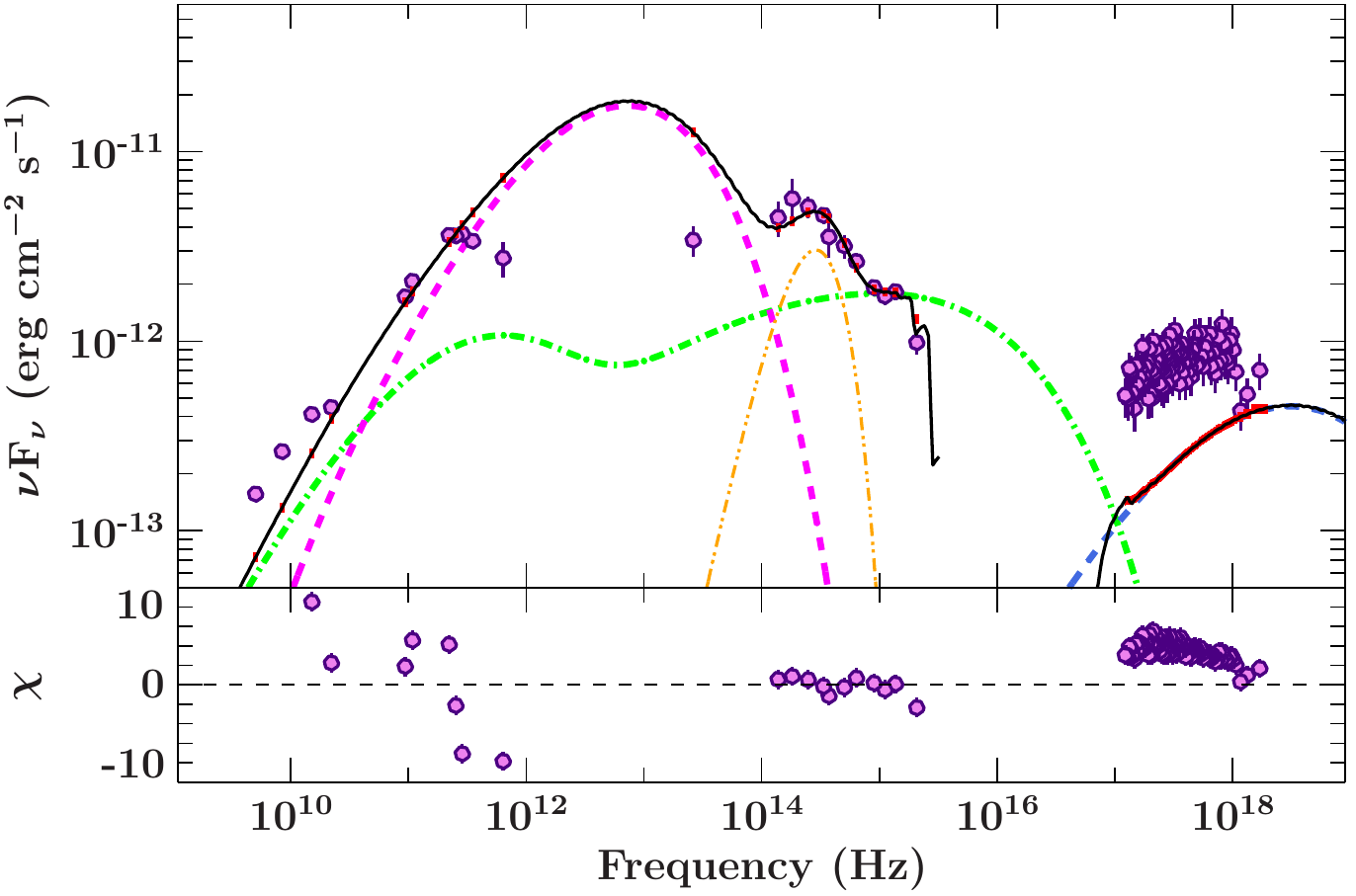}
\caption{
Radio through X-ray SED of M87 with an attempt at a SSC-dominated fit. The dashed pink line represents thermal synchrotron emission from the jet base; the dot-dashed green line de-absorbed, non-thermal synchrotron emission, the double dot-dashed yellow line the black body contribution; the dashed blue line the inverse-Compton emission. The black solid line shows the total model flux and includes the effects of absorption. The data cannot be reproduced with an SSC-dominated inner jet.
\label{fig-sedbljet2}
}
\end{figure}

Unlike in the synchrotron-dominated regime, we could not find a satisfactory fit to the data in a regime in which the X-rays are produced through synchrotron-self-Compton (SSC) near the base of the jets. This is because of the combination of constraints imposed by direct imaging of the jet collimation profile, combined with the main assumption underlying \texttt{bljet} (while the jet is accelerating, it is magnetically dominated). A highly magnetized base for a given synchrotron luminosity (fixed by the radio/sub-mm fluxes) implies a low lepton number density, which in turn results in a suppression of the SSC flux. The only way to offset such an offset is to assume a much higher temperature ($\langle\gamma_{\mathrm{e}}\rangle \approx 100$) for the radiating particles in the jet base. Our best attempt to fit the data in such a regime is shown in figure \ref{fig-sedbljet2}; as shown in the figure, such a high temperature causes the nozzle's emission to vastly exceed the sub/mm and infrared data, while still not successfully matching the X-ray flux of the source. Therefore, we rule out SSC from the magnetically dominated inner jet spine as the radiative mechanism responsible for the emission detected by \texttt{Chandra}.

% ---------------------------------------------------
% ---------------------------------------------------

\section{Discussion}

The first result emerging from our combined imaging/spectral fit is that the 3FGL spectrum can not be matched by the model for the compact jet; the predicted $\gamma$-ray flux of the core is only $\approx 10^{-14}\,$erg\,cm$^{-2}$\,s$^{-1}$, 2-3 orders of magnitude below the data. The main contribution to the core's limited $\gamma$-ray flux is due to inverse Compton scattering of the host galaxy's starlight, rather than SSC. This conclusion is mainly driven by matching our model's jet dynamics and shape with those inferred from direct imaging of the outflow through VLBI.

The second result of the fit is  that the location of particle acceleration occurs very close to the black hole ($z_{\mathrm{diss}} = 97^{+13\,R_{\mathrm{g}}}_{-3\,R_{\mathrm{g}}}$); such a distance is far closer to the central engine than the acceleration distance $z_{\mathrm{acc}} = 2.5\cdot 10^{5}\,R_{\mathrm{g}}$ inferred from the jet speed and collimation profile. Interestingly, high-resolution VLBI 86 GHz images of the jet show a ``pinching'' of the outflow around this distance \citep{Hada16}, which were also observed at this scale in GRMHD simulations (e.g.\citealt{McKinney06}, \citealt{BarniolDuran17}, \citealt{Nakamura18}, \citealt{Chatterjee19}); we tentatively suggest that the initial injection of particle acceleration in the jet may be influenced by this pinching region. 

The third result is that, assuming that the magnetically dominated jet creates most of the observed X-rays, then the radiating leptons need to be accelerated to very high Lorentz factors ($\approx 10^{7}-10^{8}$, varying slightly along the length of the jet) in order to extend the synchrotron spectrum up to the \Chandra energy range. Such high particle energies can be achieved by assuming a very high particle acceleration efficiency $f_{\mathrm{sc}}$. \cite{Prieto16} modelled a similar SED with \texttt{agnjet} but instead found a matter-dominated jet base, in which the soft X-ray photons are produced by SSC emission from the jet nozzle. We can not reproduce such a solution because in \texttt{bljet} the base of the jet is always magnetically-dominated, thus suppressing the SSC flux; this assumption leaves non-thermal synchrotron as the only radiative mechanism in the jet capable of matching the X-ray data. 

The fourth result is that the particle distribution in the jet is consistent with being isothermal even beyond the dissipation region (in our model this corresponds to $f_{\mathrm{heat}}= 1$), and the temperature of the relativistic Maxwellian ($\gamma_{\mathrm{e}}=4.2^{+1.0}_{-0.3}$) is well constrained by the sub-mm ALMA data, which forms a clear bump. The finding that the particle distribution in the jet is isothermal differs from our findings in Paper \RN{1}, in which we showed that it was necessary to \textit{increase} the temperature ($f_{\mathrm{heat}}\gg 1$) of the particle distribution at $z_{\mathrm{diss}}$ (figure 12 in Paper \RN{1}). One likely explanation for this discrepancy is in the difference in jet kinetic powers between the two sources. In PKS\,2155$-$304 the higher kinetic power ($N_{\mathrm{j}}=9^{+0.6}_{-0.7}\cdot10^{-3}\,L_{\mathrm{Edd}}$) could drive stronger shocks in the jet, allowing for additional energy to be transferred to the radiating particles, while in M87 ($N_{\mathrm{j}}=3.2^{+0.5}_{-0.3}\cdot10^{-5}\,L_{\mathrm{Edd}}$) this amplification does not seem to be necessary.

The optical/IR thermal bump was interpreted by \cite{Prieto16} as a tracer of an optically thick, geometrically thick accretion disk, with an inner radius of $5\,R_{\mathrm{g}}$ and temperature of $3200\,\mathrm{K}$; when fitting a Shakura-Sunyaev disk to the data we recover similar parameters. However, this combination of inner radius and temperature results in an accretion rate of $\approx 10^{-7}\,\mathrm{\dot{M}_{Edd}}$. At such a low accretion rate the disk should be in the ADAF state, and therefore its emission should not resemble a black body (or a superposition of black bodies). Furthermore, such low accretion rate is two orders of magnitude below the estimated jet power. These findings suggest that the origin of the thermal bump may not be related to the accretion disk, and instead be caused by a residual starlight contribution in the inner 32 pc of the galaxy. 

The transition in the jet shape from parabolic to conical, as well as the observed trend of increasing bulk motion up to HST-1, imply that in M87 bulk acceleration continues up to large scales of $\approx 10^{5}\,{R_\mathrm{g}}$. Assuming that the process responsible for accelerating the jet is the dissipation of magnetic field into bulk kinetic energy in a highly magnetized region, this implies that the outflow remains magnetically-dominated ($\sigma \geq 1$) up to these large scales; high $\sigma$ in turn implies a relatively low lepton number density required to match the synchrotron spectrum, and such low lepton number density naturally results in a suppression of the inverse Compton flux. The inefficiency of the Inverse Compton process in the source is unchanged by taking lower values of $\sigma_{\mathrm{acc}}$: the regions near the dissipation region $z_{\mathrm{diss}}$ (responsible for the radio and X-ray emission) are always highly magnetized, thus automatically setting a relatively low number density of particles throughout the outflow. Increasing the Inverse Compton flux would require both taking a low $\sigma_{\mathrm{acc}}$ and increasing the jet power, which would in turn cause the synchrotron emission to exceed the radio/sub-mm data. Our findings are in contrast to previous modelling efforts of M87 (e.g. \citealt{Abdo09}, \citealt{deJong15}), who reproduced the SED of the source with a standard homogeneous one-zone SSC model. The parameters explored in both of these works require the plasma emitting in the core region to be strongly particle-dominated, with $U_{\mathrm{e}}/U_{\mathrm{b}} \geq 100$, in order to produce a meaningful high-energy flux; this is in contrast with VLBI data, which favours a magnetically-dominated core ($10^{-1} < U_{\mathrm{e}}/U_{\mathrm{b}} < 10^{-4}$, \citealt{Kino14}, \citealt{Hada16}). 

Recently, coupling radiation with GRMHD simulations has also been used to model the inner jet of M87 in place of simple semi-analytic models (e.g. \citealt{Moscibrodzka16}, \citealt{Chael18}). Both of these works find that the bulk of the X-ray emission is due to SSC rather than optically thin synchrotron, in contrast to our work here. However, this conclusion depends very strongly on the assumptions made in the post-processing of the radiation in the simulations due to two main factors. Firstly, in GRMHD simulations the radiating electrons are assumed to predominantly be located in the outer sheath of the jet, thus restricting the emitting region to a more matter-dominated region than the jet base of our model. Secondly, the particle distribution in both of the above works is assumed to be only a relativistic Maxwellian, which prevents the optically thin synchrotron spectrum from extending above the sub-mm band. Finally, it is worth noting that if SSC were the dominant mechanism producing X-rays in M87, it would no longer be consistent with a low power blazar when oriented face-on (see section 4.2). Because of these differences, a direct comparison of the spectra predicted from these simulations and from our model is not straightforward.  Including more realistic particle distributions which account for non-thermal leptons \citep{Davelaar18} in simulations will facilitate such a comparison in the future. 

\cite{EHT19.5} used the X-ray flux as a constrain of several GRMHD models by (conservatively) rejecting all the solutions whose SSC emission over-predicts the data. Our work suggests instead that more GRMHD model could be rejected by the observations, as we expect the bulk of the X-ray radiation to originate from non-thermal synchrotron emission.

\subsection{The origin of the $\gamma$-ray emission}

Our work shows for the first time that in the context of a leptonic model based on an MHD-driven, magnetically accelerated outflow, the $\gamma$-ray emission of M87 likely does not originate in the magnetically dominated inner jet regions. In this section we explore alternative mechanisms for the high-energy emission, and discuss the implications of each.

One possible way of increasing the efficiency of IC is if the jet is structured, with an inner, fast spine surrounding a slow-moving sheath, as commonly found in GRMHD simulations, e.g. \cite{McKinney06}, \cite{Hardee07}, \cite{Penna13}, \cite{Nakamura18}, \cite{Chatterjee19}. In this case, the synchrotron emission from the spine/layer is boosted in the co-moving frame of the sheath/spine, resulting in an increase of the IC flux  \citep{Ghisellini05}. Indeed, such a model was applied to M87 by \cite{Tavecchio08}, who showed that the TeV emission could be well matched thanks to the spine/sheath contribution to the seed photon fields. However, an additional Inverse Compton contribution would reduce the radiative cooling time-scale of high energy particles, which are needed to match the \Chandra%and \Nu
data, thus requiring extremely short acceleration time-scales. A similar issue was also pointed out by \cite{Costamante18} for the case of hard-TeV BL Lacs. 

Alternatively, the high-energy bump could be due to hadronic processes. The main caveat to this scenario is the high power requirement typically associated with hadronic models. Our estimated jet power, driven mainly by the radio fluxes, is $3,2\cdot 10^{-5}\,L_{\mathrm{Edd}} = 2.9\cdot 10^{43}\,\mathrm{erg s^{-1}}$ assuming a $6.5\cdot 10^{9}\,\mathrm{M_{\odot}}$ black hole. This power assumes one cold proton (carrying the jet's bulk kinetic energy) per electron. Such an estimate is consistent with, but on the lower end of, independent measures either studying the internal pressure exerted by the kpc-scale jet knots (e.g. \citealt{Bicknell96}, \citealt{Owen00}, \citealt{Stawarz06}), or by estimating the required heating in the galaxy's X-ray halo (e.g. \citealt{Churazov02}, \citealt{Forman05}, \citealt{Allen06}, \citealt{Russell13}); all of these find a range of $P_{\mathrm{jet}}\approx 10^{-5}-10^{-4}\,L_{\mathrm{Edd}}$. Requiring that the power in the halo/outer jet and core be roughly of the same order of magnitude, these constraints on the jet's energetics leave relatively little room for a population of relativistic protons in the outflow (unless the core has recently entered a phase of renewed activity). However, implementing an energetically dominant population of relativistic/hot particles (either protons or leptons) would cause the underlying assumptions of \texttt{bljet} to fail (see Paper \RN{1}), and is therefore beyond the scope of this work. 

A third possibility is that the high-energy emission does not originate in the core, but in the kpc-scale jet (\citealt{Stawarz03}, \citealt{Hardcastle11}), due to a combination of IC with synchrotron, stellar and cosmic microwave background photons. The recent hints of variability on monthly time scales found in the \Fermi/LAT light curve of the source \citep{Ait18} as well as the fast TeV variability \citep{Aharonian06} would disfavour such an interpretation. \cite{Cheung07} however showed that the inferred size of the HST-1 complex at VLBI scales is compact enough that it could indeed be the source of the TeV emission; furthermore, flaring activity on yearly time scales has been detected in X-ray kpc-scale jets of Pictor A \citep{Marshall10} as well as M87 itself \citep{Harris03}. Because of all of these arguments, we conclude that HST-1 or the kpc-scale jet can not be ruled out as the sites of a significant portion of M87's $\gamma$-ray emission. 

In addition to the limited variability, \cite{Ait18} found that the source shows a complex spectrum likely originating from multiple components, of which at least one is variable. Such complex behaviour could be reproduced if, on top of a steady state component (such as the large scale jet), a secondary highly variable region is also present. One possible candidate in such scenario would be magnetospheric gap acceleration in the black hole's ergosphere (e.g. \citealt{Neronov07}, \citealt{Rieger08}, \citealt{Levinson11}, \citealt{Moscibrodzka11}, \citealt{Broderick15}).

In conclusion, while the location of the $\gamma$-ray emission is still unclear, our work rules out a one zone SSC model originating in the magnetized core, as such mechanism implies plasma conditions in strong disagreement with theoretical expectations.

\begin{table}
\begin{tabular}{lll}
\hline
Source name & Redshift  & Source classification\\ 
\hline
5BZGJ0048+3157 & 0.015 & BL Lac\\ 
5BZGJ0153+7115 & 0.022 & BL Lac\\ 
5BZGJ0204+4005 & 0.007 & BL Lac\\ 
5BZUJ0241-0815 & 0.005 & Blazar Uncertain type	\\ 
5BZUJ0319+4130 & 0.018 & Blazar Uncertain type	\\ 
5BZGJ0709+501 & 0.02 & BL Lac\\ 
5BZGJ1148+592 & 0.011 & BL Lac\\ 
5BZUJ1301-3226 & 0.017 & Blazar Uncertain type	\\ 
5BZUJ1325-4301 & 0.002 & Radio Galaxy: Cen A\\ 
5BZGJ1336-0829 & 0.023 & BL Lac\\ 
5BZGJ1407-2701 & 0.022 & BL Lac\\ 
5BZUJ1632+8232 & 0.025 & Blazar Uncertain type	\\ 
5BZGJ1719+4858 & 0.024 & BL Lac\\ 
5BZGJ1840-7709 & 0.018 & BL Lac\\ 
5BZGJ1945-5520 & 0.015 & BL Lac\\ 
5BZUJ2209-4710 & 0.006 & Blazar Uncertain type	\\ 
\hline
\end{tabular}
\caption{List of our comparison sample from the ROMABZCAT catalogue, along with their redshift and classification. We excluded Centaurus A from our comparison as it is not seen face-on.}
\label{tab-blazars}
\end{table}

\subsection{What kind of misaligned blazar is M87?}

\begin{figure}
%\hspace{-0.5cm}
\includegraphics[scale=0.58]{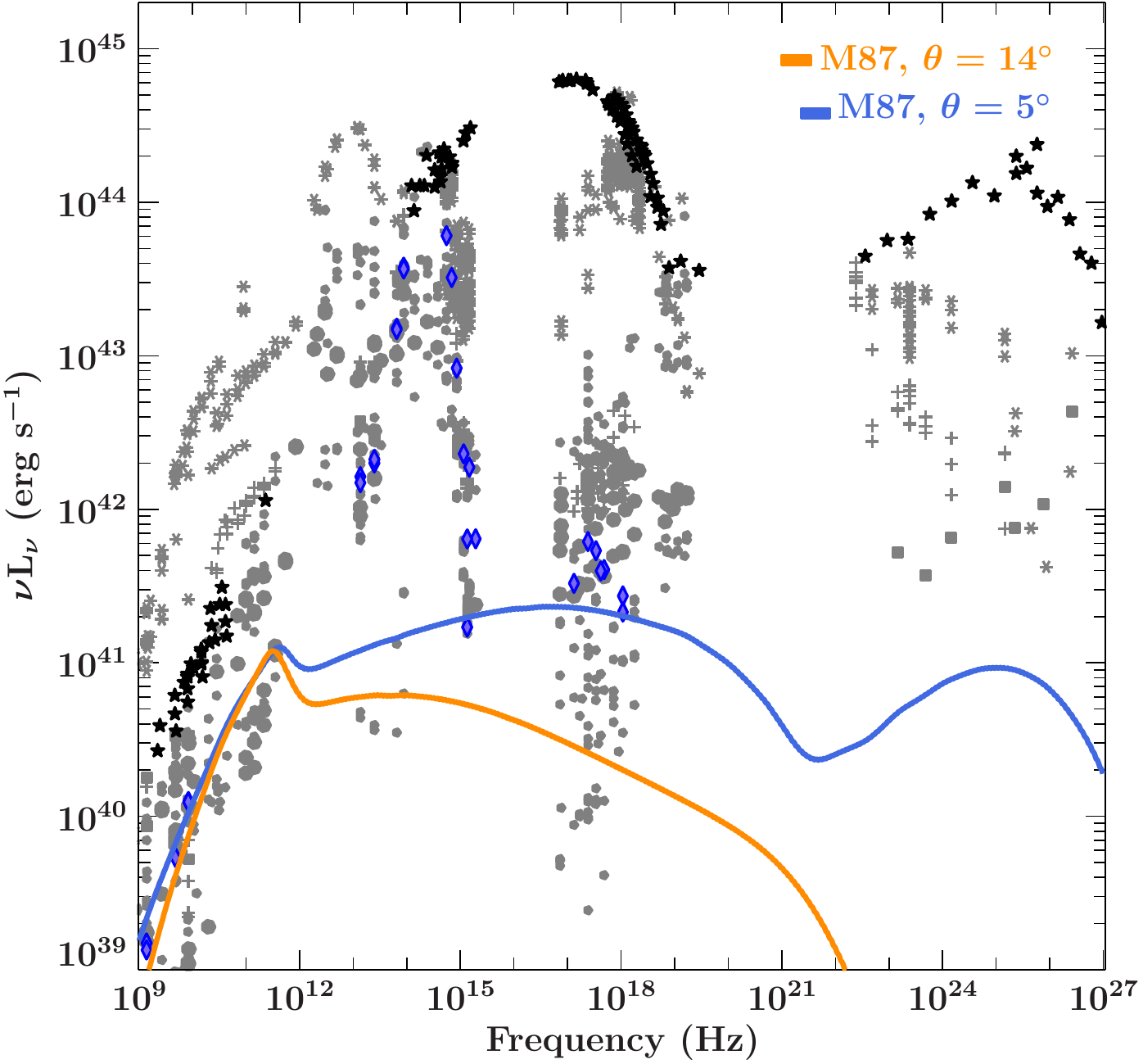}
\caption{
Comparison between our model for M87 (orange line), the same model but re-scaled to a viewing angle of $5^{\circ}$, our sample of nearby blazars (grey points), and Mrk 421 (black stars). The \Fermi/LAT-detect sources are represented by squares (5BZGJ0153+7115), asterisks (5BZUJ0319+4130), and plusses (5BZUJ1632+8232) respectively. The SED of 5BZGJ0709+501 is shown by the blue diamonds. The SED of M87 seen face-on is clearly far more faint than that of $\gamma$-ray bright HBLs. Out of all sources in our sample, we found that the most similar blazar SED to our aligned model is that of 5BZGJ0709+501.
}
\label{fig-M87blazar}
\end{figure}

Figure \ref{fig-M87blazar} shows a comparison between our best fitting model re-scaled to a viewing angle of $5^{\circ}$, and a sample of SEDs from nearby blazars. We built the sample by selecting all blazars in the ROMABZCAT catalogue \citep{Massaro15} with known redshift between 0 and 0.025, corresponding to a luminosity distance of about 110 Mpc. This search returned 16 sources, listed in table \ref{tab-blazars}. One of these sources (5BZUJ1325-4301) is the misaligned radio galaxy Centaurus A, and thus we excluded it from the sample. Out of the remaining 15 sources, 10 are listed as galaxy-dominated BL Lacs and 5 as blazars of uncertain type; 3 are detected by \Fermi/LAT (5BZGJ0153+7115, 5BZUJ0319+4130, 5BZUJ1632+8232) and two are detected in radio, infrared and optical, but not in X-rays or $\gamma$-rays (5BZGJ1148+592, 5BZGJ1945-5520). The data of this sample are from \cite{Myers03}, \cite{Healey01}, \cite{Dixon70}, \cite{Jackson07}, \cite{Condon98}, \cite{White97}, \cite{Gregory91}, \cite{Nieppola07}, \cite{White92}, \cite{Kuehr81}, \cite{McConnell12}, \cite{Wright90}, \cite{Mauch03}, \cite{Wright94}, \cite{Murphy10}, \cite{Wright09}, \cite{Moshir90}, \cite{Iras94}, \cite{Gregory96}, \cite{Planck11}, \cite{Planck14}, \cite{Planck15}, \cite{Wright10}, \cite{Bianchi11}, \cite{Warwick81}, \cite{Levine84}, \cite{Evans14}, \cite{Rosen15}, \cite{Saxton08}, \cite{Voges99}, \cite{Boller16}, \cite{Elvis92}, \cite{Evans10}, \cite{Forman78}, \cite{Verrecchia07}, \cite{Hiroi11}, \cite{Hiroi13}, \cite{DElia13}, \cite{Cusumano10a}, \cite{Cusumano10b}, \cite{Ajello12}, \cite{Bird10}, \cite{Baumgartner13}, \cite{Piccinotti82}, \cite{Hartman99}, \cite{Acero15}, \cite{Abdo10}, \cite{Nolan12}, \cite{Giommi12}, and \cite{Bartoli13}. Finally, we included the averaged SED of Mrk 421 from \cite{Abdo11} to compare our model to a prototypical high-peaked BL Lac (HBL). 

% ---------------------------------------------------
% ---------------------------------------------------

When re-scaled to a face-on geometry, the SED predicted by our model shifts to qualitatively resemble a BL Lac: the increase in beaming in the outer jet regions causes the synchrotron peak to shift from the far-IR to the X-ray band, and the Compton peak luminosity almost matches the synchrotron luminosity. Despite this increase in beaming, the SED remains more faint by three orders of magnitude than Mrk 421, and is one to three orders of magnitude more faint than the other \Fermi-detected sources in our sample. Instead the predicted SED resembles the more faint, galaxy-dominated sources of the sample, which are detected in radio and X-ray surveys but not at higher energies. In particular, our face-on model is in remarkably good agreement with the radio and X-ray data of 5BZGJ0709+501.

The reason for the low luminosity is the available energy budget in the jet inferred from our model, which is far lower than that we found for PKS$\,$2155$-$304 ($P_{\mathrm{jet, M87}} \approx 3\cdot 10^{-5}\,L_{\mathrm{Edd}} = 2.9\cdot 10^{43}\,$erg\,s$^{-1}$ against $P_{\mathrm{jet}, 2155} \approx 9\cdot 10^{-3}\,L_{\mathrm{Edd}} = 1.25\cdot 10^{45}\,$erg\,s$^{-1}$, see Paper \RN{1}). This finding implies that M87 is not the misaligned counterpart of a typical $\gamma$-ray bright HBL like Mrk 421 or PKS$\,$2155$-$304, which are inherently more powerful sources.

Despite the low energy budget, radio images of M87 display a typical FR\RN{1} morphology: the jet extends for tens of kilo-parsecs and terminates in well-developed lobes \citep{Owen00}. The ability of a low-power jet to produce such extended structure suggests that the surrounding environment of the galaxy, rather than the central engine, may be the main driver for the formation of the large scale radio structure (e.g. \citealt{Tchekhovskoy16}).

\section{Conclusion}
In this paper we have for the first time combined constraints from VLBI imaging and spectral data to model the jet from the M87 radio galaxy, using our multi-zone jet model, \texttt{bljet}. We find that \texttt{bljet} can reproduce both the jet morphology and the SED from the radio to the hard X-ray band. Furthermore, the strong constraints imposed by the data ensure that little degeneracy in the model is present, and that the free parameters are well-determined. In particular, we find that the jet power near the core is $P_{\mathrm{jet, M87}} \approx 3.2 \cdot 10^{-5}\,L_{\mathrm{Edd}} = 2.9\cdot 10^{43}\,$erg\,s$^{-1}$, in agreement with previous independent estimates. Deriving such good constraints highlights the importance of combining spectral modelling with additional information such as direct VLBI imaging when modelling AGN jets.

In the $\gamma$-ray regime we find that the inner magnetically-dominated jet predicts too little flux to reproduce the \Fermi/LAT 3FGL spectrum. This is because the high magnetization at the base, combined with the radio flux constraints, implies a low lepton number density, resulting in a very low inverse Compton flux. Our findings are in contrast with single-zone SSC models, which can match the high-energy SED but only if the plasma is highly matter-dominated (e.g. \citealt{Abdo09}, \citealt{deJong15}). Such a strongly matter-dominated emitting region is unlikely to exist in the magnetically-dominated inner jet, but might exist in the outer sheath of the jet \citep{Tavecchio08}. Additional mechanisms that could allow for enhanced $\gamma$-ray emission are particle acceleration in the vicinity of the black hole (either the ergosphere, e.g. \citealt{Neronov07}, \citealt{Rieger08}, \citealt{Levinson11}, or the stagnation surface, \citealt{Broderick15}), or inverse-Compton scattering of the host galaxy's starlight and/or of the CMB in the large scale jet \citep{Stawarz03}. 

When re-scaling our best-fit model to a face-on  viewing angle of $5^{\circ}$, we find that while the SED roughly resembles that of a BL Lac, the predicted luminosity remains far lower than a prototypical HBL like MrK 421. The main reason is that the increased beaming does not offset the low available jet power. This finding implies that the jet of M87 is not powerful enough to be the misaligned counterpart of a $\gamma$-ray bright BL Lac, instead resembling more a faint, galaxy-dominated source.

The main caveat of our model is that it focuses on probing the inner magnetically-dominated spine, neglecting the emission from the (more particle-dominated) outer jet sheath. On the one hand, limb-brightening is clearly observed in radio imaging of M87 (e.g. \citealt{Mertens16}, \citealt{Hada16}), and in principle the sheath would be a more efficient inverse Compton emitter than the spine, as it is more particle-dominated than the latter. The additional radiation from the sheath could lead to enhanced X-ray or even $\gamma$-ray emission, which in turn would affect our conclusions. On the other hand, our decision to focus on the spine is supported by further arguments, both theoretical and observational. First, the inner spine determines the shape and size of the sheath/interface region with the outer disk (e.g. \citealt{McKinney06}, \citealt{Tchekhovskoy10}, \citealt{Nakamura18}, \citealt{Chatterjee19}), implying that our assumed shape for the spine is consistent with the observed shape of the (edge-brightened) jet. Second, our synchrotron-dominated core model, when re-scaled to a face-on geometry, resembles a low power BL Lac, in agreement with AGN unification models (\citealt{Antonucci93}, \citealt{Urry95}). This would not be the case if the bulk of the X-rays were produced in the slower outer sheath, as the Doppler factor would not be expected to vary significantly between the two viewing angles. Therefore, while our model cannot fully capture the physics of the system, it seems to provide a good overall description of the source beyond simpler single-zone models.

Our results are particularly important in light of the upcoming observations of M87 with the Event Horizon Telescope (e.g. \citealt{EHT19.1}), which provide even more detailed imaging of the regions near the black hole. These state-of-the-art observations will require further improvements in the modelling of jets, for example by explicitly solving the relativistic MHD equations in the presence of gravity (e.g. \citealt{Polko10}, \citeyear{Polko13}, \citeyear{Polko14}, \citealt{Ceccobello18}) to derive the jet dynamics. We plan to couple these more physically consistent  MHD solutions to observables in future works.

\section*{Acknowledgements}
We are grateful to Masanori Nakamura, Keiichi Asada and Kazuhiro Hada for providing us with their VLBI measurements of the jet profile in M87. M. L. and S. M. are thankful for support from an NWO (Netherlands Organisation for Scientific Research) VICI award, grant Nr. 639.043.513. F. K. acknowledges funding from the WARP program of the Netherlands Organisation for Scientific Research (NWO) under grant agreement Nr: 648.003.002. This research has made use of \ISIS functions ({\ISIS}scripts) provided by  ECAP/Remeis observatory and MIT (http://www.sternwarte.uni-erlangen.de/isis/). Part of this work is also based on archival data and online services provided by the ASI Science Data Center(ASDC). This publication makes use of data products from the Wide-field Infrared Survey Explorer, which is a joint project of the University of  California, Los Angeles, and the Jet Propulsion Laboratory/California Institute of Technology,funded by the National Aeronautics and Space Administration.

\end{document}